\documentclass[aps,pra,11pt,reprint,floatfix]{revtex4-1}

\usepackage{lineno}
\usepackage{amssymb}
\usepackage{amsmath}
\usepackage{graphicx}
\usepackage{hyperref}

\DeclareMathOperator*{\argmax}{arg\,max}

\begin{document}

\title{Near-field speckle-scanning-based x-ray imaging}
\author{Sebastien Berujon}
\email[]{berujon@esrf.eu}
\affiliation{European Synchrotron Radiation Facility, CS40220, 38043 Grenoble Cedex 9, France}
\author{Eric Ziegler}
\affiliation{European Synchrotron Radiation Facility, CS40220, 38043 Grenoble Cedex 9, France}

\date{\today}

\begin{abstract}
The x-ray near-field speckle-scanning concept is an approach recently introduced to obtain absorption, phase, and dark-field images of a sample. In this paper, we present ways of recovering from a sample its ultrasmall-angle x-ray scattering distribution using numerical deconvolution. We also show how to access the 2D phase gradient signal from random step scans, the latter having the potential to elude the flat-field correction error. Each feature is explained theoretically and demonstrated experimentally at a synchrotron x-ray facility.
\end{abstract}
\maketitle
\section{Introduction}

When X-ray imaging was discovered, it took only a few months to the scientific community to express a great interest, as acknowledged by the award of the first Nobel Prize to W. R\"ontgen. Since then fast and tremendous progress could be observed in diverse applications over domains including medicine, material science, paleontology and many others. Within more than a century great minds realized the importance of X-ray imaging and contributed to its development with, for instance, the introduction of the dose concept \cite{cantril1945,parker1950} and the advent of computed tomography \cite{chiro2979,hounsfield1973}. In parallel, constant progress in instrumentation allowed the development of X-ray imaging systems with ever better contrast, resolution and efficiency \cite{kb1948,bonse1965,momose1996}.

While questing for means of collecting more information and reducing the dose necessary to image a specimen, the appearance of the first coherent X-ray sources favoured the emergence of phase contrast imaging methods \cite{bonse1965,snigirev1995,cloetens1996}. Indeed, these were motivated by the fact that, in the X-ray domain, the material optical index $n = 1-\delta -i\beta$ presents usually a refractive index $\delta$ with a magnitude several orders larger than its absorption counterpart $\beta$ \cite{henke1993}. Thus, since the inception of coherent third generation synchrotron sources, different near-field techniques of X-ray phase sensing emerged, either inspired by their analogue in other spectral domains or specially developed \cite{momose2005}. One may categorize those imaging techniques depending whether they are sensitive to the Laplacian or the gradient of the phase. The propagation-based methods, which belong to the first category, make use of the property of contrast enhancement of edges when increasing the distance to the detector from the specimen under study \cite{nugent1996,cloetens1999,paganin2004,langer2008}. The second category brings together intruments such as deflection-based instruments, including X-ray grating interferometers \cite{weitkamp2005,momose2006}, Hartmann like sensors \cite{mercere2005,rizzi2013}, coded apertures \cite{olivo2007,munro2012}, analyser based systems \cite{davis1995} and speckle based methods \cite{berujonPRL,morgan2012}. Other full-field imaging approaches available in the X-ray domain and using different principles include for instance ptychography \cite{stockmar2013} and Zernike phase microscopy \cite{holzner2010,yang2014}.

Propagation based methods offer the advantage of a high sensitivity to sharp changes in electronic density, the sample high frequency features being rendered with high contrast thanks to the phenomenon of optical interference. Conversely, they often suffer from reconstruction artefacts when imaging materials presenting a slowly varying density and low spatial frequencies due to the very nature of the recorded signal, i.e. a Fresnel diffraction image \cite{diemoz2012}. In practice, a phase gradient sensitive method would be preferable to image a homogenous material, although usually resulting in a spatial resolution lower than the one obtained with a propagation based technique, all other conditions remaining equal. For their implementation, phase-gradient sensitive methods require the use of a specific optics (grid, grating, crystal...) to modulate the wavefront, which can, sometimes, be expensive or difficult to get.

Darkfield X-ray contrast is an additional contrast mode inspired from its counterpart in the visible light and made possible thanks to the availability of crystal analysers \cite{davis1995,miles2003} and of grating interferometers \cite{pfeiffer2008}. Darkfield imaging permits to reveal the structure and the orientation of features having a characteristic size much smaller than the detector resolution. For instance, the mapping of the dark field signal with a grating interferometer allows one to determine the local reduction of the interference amplitude caused by X-ray beam propagation through the sample. The interference amplitude actually reflects the progressive degradation of the beam transverse coherence properties caused by scattering induced upon photon propagation through the samples sub-microscopic features.

The X-ray near field speckle scanning technique \cite{berujonPRA,berujonPhD} is a recent and particularly attractive deflection based technique as it provides both phase and darkfield contrast images in addition to the traditional absorption one. Moreover, it uses only a random phase object as wavefront modulator. This phase object is used to generate a speckle intensity pattern in the near field domain caused by mutual interference of the scattered waves. The near-field domain is particularly interesting because the speckle pattern distortion is only ruled by the wavefront transformation upon propagation in the Fresnel zone \cite{gatti2008,magatti2009}, a region easily accessible with X-rays \cite{cerbino2008}.

The X-ray near field speckle tracking technique previously developed and easily implemented \cite{berujonPRL,morgan2012} could provide phase information from one single exposure of the sample. However, its resolution is limited by the speckle grain size and artefacts may appear in the case of complex objects. The speckle scanning method appeared to overcome these limitations, nonetheless at the cost of additional image acquisitions. The technique advantages include a high sensitivity to the phase gradient, a moderate requirement for the beam monochromaticity, a high photon efficiency (no absorption in the system) and the use of a minimalist optical element in the form of a scattering membrane. Albeit the speckle based methods were first developed at a synchrotron source, they were recently adapted with success to a laboratory source, thereby broadening the range of applications ~\cite{zanette2014}.
In this letter, several new aspects of the speckle scanning concept are presented. After having recalled the basic concept and the general experimental setup, we will expound a method capable of accessing the ultra-small angle X-ray scattering distribution of a sample from 2D rasters scans. Next, a processing scheme to recover the phase gradient information in two dimensions from scans with random steps is demonstrated. Each aspect is illustrated with experimental applications. We also describe a way of eluding the 'flat-field correction error' using the speckle scanning technique. A short discussion dealing with the sensitivity and operational constraints is concluding this paper.

\section{Experimental setup}
\begin{figure}
\centerline{\includegraphics[width=8cm]{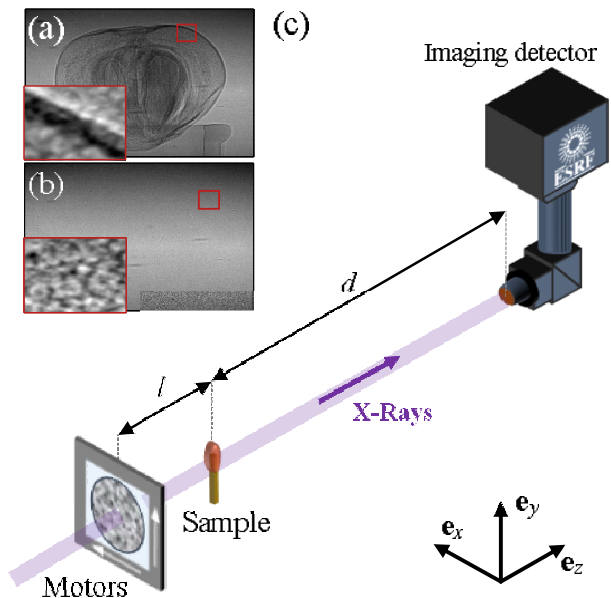}}
\caption{(Color online) First image recorded during the scan with (a) and without an object (b) inserted in the beam. (c) Experimental setup for the data collection.}\label{fig:setup}
\end{figure}

Let us introduce the experimental setup that was used to acquire the data presented further. The experiments were conducted at the beamline BM05 of the ESRF \cite{ziegler2004}. There, the photons are produced by synchrotron radiation from a bending magnet of 0.85~T installed on a storage ring operating with 6.04~GeV electrons. In the experiments presented, a beam with an energy E = 17~keV was selected thanks to a double crystal Si(111) monochromator with a selectivity of $\Delta E/E \sim 10^{-4} $. The experimental setup sketched in Fig.~\ref{fig:setup} was installed in a lead-shielded hutch, at distance from the scattering object to the source of $R\approx55$ m. Considering the source size of 29~$\mu$m rms vertically and $\sim$77~$\mu$m rms horizontally, the transverse coherence at the level of the membrane at E = 17~keV is expected to be at most $\sim$29~$\mu$m vertically by 11~$\mu$m horizontally.

The scattering object, a piece of sand paper with a grit designation P800, was mounted on a motorized 2D piezo stage, itself mounted above stepping micromotors. Such a setup permitted to move the random phase object transversally to the beam with either a nanometer precision and a 100 $\mu$m range, or a micrometer accuracy over a millimeter range. The scatterer was placed at a distance $l = 450$ mm upstream of the sample and the detector at a variable distance $d$ downstream from it. The imaging system was a FReLoN (Fast Read-Out Low Noise) CCD camera receiving a visible light signal resulting from the luminescence produced by X-ray illumination of a scintillator and imaged by a microscope objective. The system effective pixel size was of $s_p = 5.8~\mu$m. Near field speckle of a size of a few pixels was observable in the recorded images with a contrast defined by a coefficient of variation of $\sim 14\%$ when $d=850$ mm (visible in Fig.~\ref{fig:setup}~(a) and (b)).

\section{Scanning technique with constant step\label{sec:backgr}}
This speckle scanning mode is mainly presented in Ref.~\cite{berujonPRA} where the concept is described as the crossroads of the speckle tracking \cite{berujonPRL,morgan2012} and grating interferometry techniques \cite{creath1988,goldberg2001}. Since we are going to present an extension of it, let us recall quickly the method principle.
\begin{figure}
\centerline{\includegraphics[width=8cm]{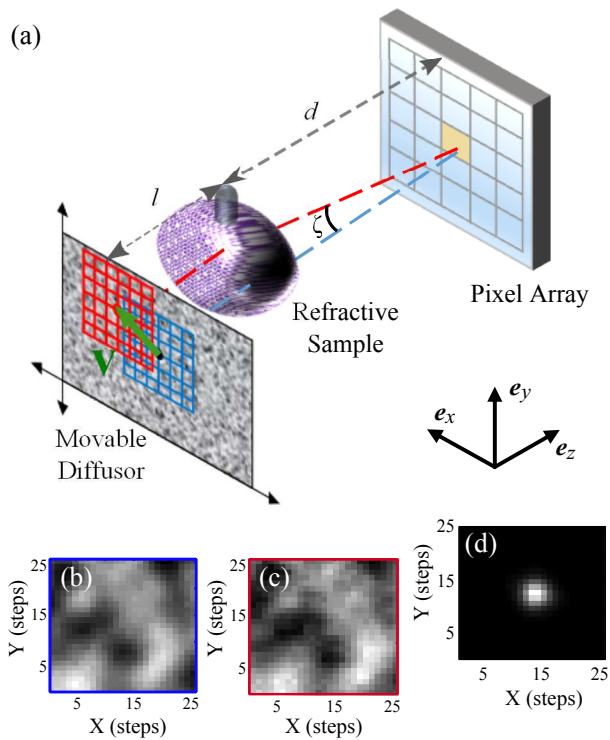}}
\caption{(Color online)(a) Geometry for the scanning X-ray speckle imaging technique with constant steps. (b) Example of a reference speckle pattern built from the intensity values read in the yellow pixel during the scan. (c) Corresponding pattern recorded in the same pixel during a scan with the sample in the beam. (d) Scattering distribution obtained by deconvolution of (b) with (c).\label{fig:NormSketch}}
\end{figure}

Figure~\ref{fig:NormSketch} shows the geometry considered wvith the vectors $(\mathbf{e_{x},e_{y}})$ perpendicular to the beam propagation direction $\mathbf{e_z}$. The scattering object with random phasors is placed in a partially coherent X-ray beam, either right upstream or downstream the investigated object. While small differences exist between these two configurations \cite{berujonAPL2013,morgan2012} they are negligible in this following experiments as the beam was nearly collimated. Nonetheless, we will rederive here the correct formulae for the magnifying geometries currently encountered, e.g. when using a focusing optics or a highly divergent laboratory X-ray source.

\begin{figure*}
\centerline{\includegraphics{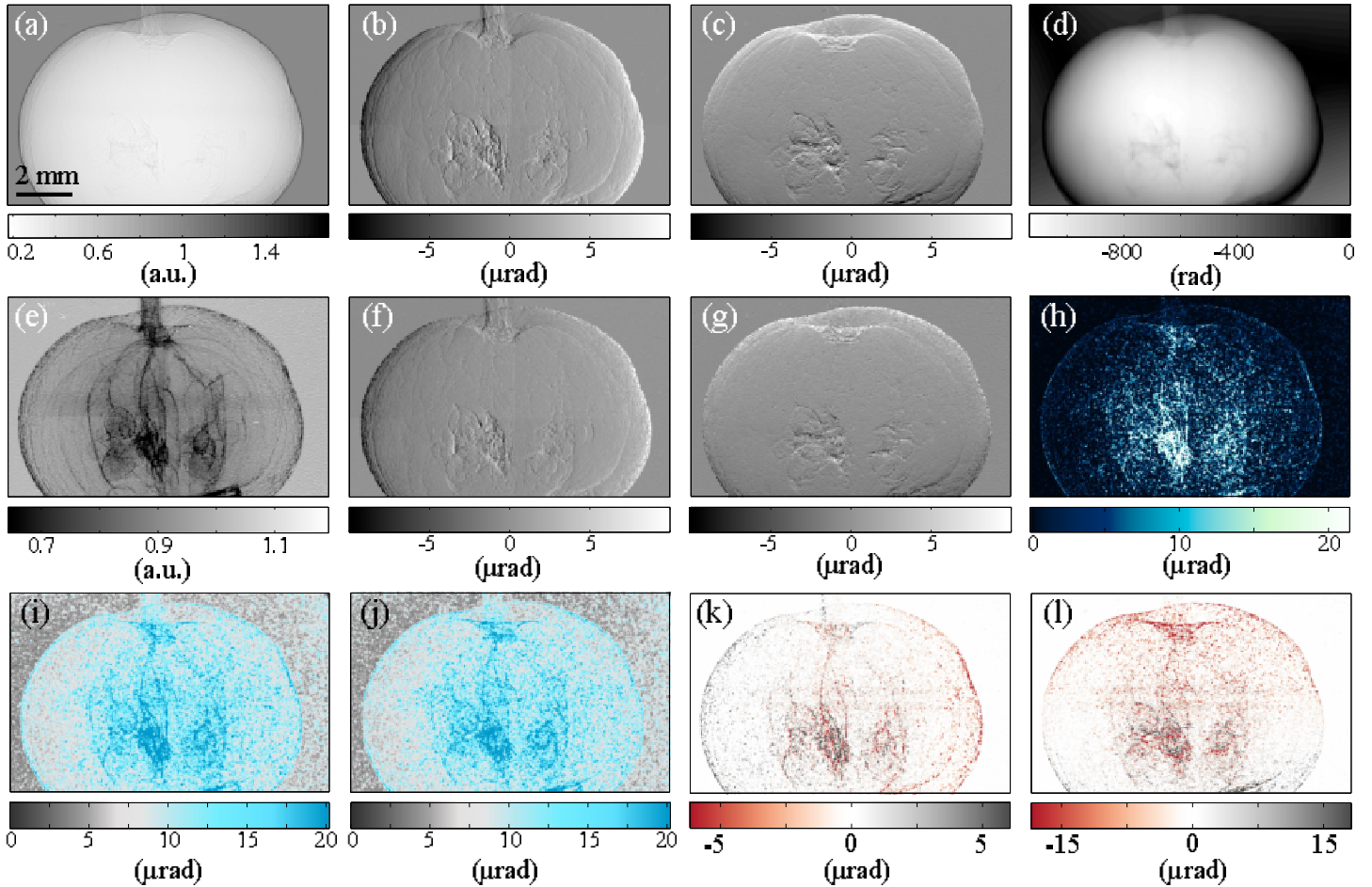}}
\caption{(Color online) Image modes of a blueberry sample obtained with the method of Sec.~\ref{sec:backgr}: (a) transmission image $T$, (b-c) horizontal and vertical wavefront gradients $\nabla W.\mathbf{e_{x/y}}$, (d) phase reconstruction $\phi$ and (e) scattering image $Df$. Images obtained using the method of Sec.~\ref{sec:usmangles}: (f-g) horizontal and vertical wavefront gradients $\nabla W.\mathbf{e_{x/y}}$. Higher moments of the scattering distribution: (i) $\sqrt{M_{20}}$,(j) $\sqrt{M_{02}}$, (k) $\sqrt[3]{M_{30}}$, (l) $\sqrt[3]{M_{03}}$ and (h) $\sqrt[4]{M_{40}}$. \label{fig:mytille}}
\end{figure*}

The method consists of coupling together data collected when raster scanning the membrane with the sample present in the X-ray beam (e.g. array of Fig.~\ref{fig:setup}.(c)) or removed from it (Fig.~\ref{fig:setup}.(b)). For the technique to be effective, the membrane must be scanned with a positional repeatability $\mathbf{p}=(X,Y)$ of a fraction of a speckle grain size and with a strictly constant step size $s_{s}= cst$. A pair of patterns recorded in a pixel located at $\mathbf{r} = (x,y)$ is considered to be independent from the data collected in the neighboring pixel. The pixel marked in yellow in Fig.~\ref{fig:NormSketch}(a) with the pattern pair displayed in (b-c) is an example of data sets recorded at a given position $\mathbf{r}$. Then the 2D intensity value array recorded when scanning the membrane with the sample in the beam path and noted $f$ (Fig.~\ref{fig:NormSketch}.(c)) is cross-correlated with the one obtained during the reference scan and noted $g$ (Fig.~\ref{fig:NormSketch}.(b)), i.e. with no sample in the beam. The outcome of the operation is a correlation map whose maximum peak location indicates the displacement vector $\mathbf{v}$ between the two arrays:
\begin{equation}
\mathbf{v}(\mathbf{r}) = \argmax_{\mathbf{\tau}} \int f(\mathbf{\mathbf{r},p}) g(\mathbf{r},\mathbf{p}+\mathbf{\tau})d\mathbf{\tau}
 \end{equation}

\begin{figure}
\centerline{\includegraphics[width=8cm]{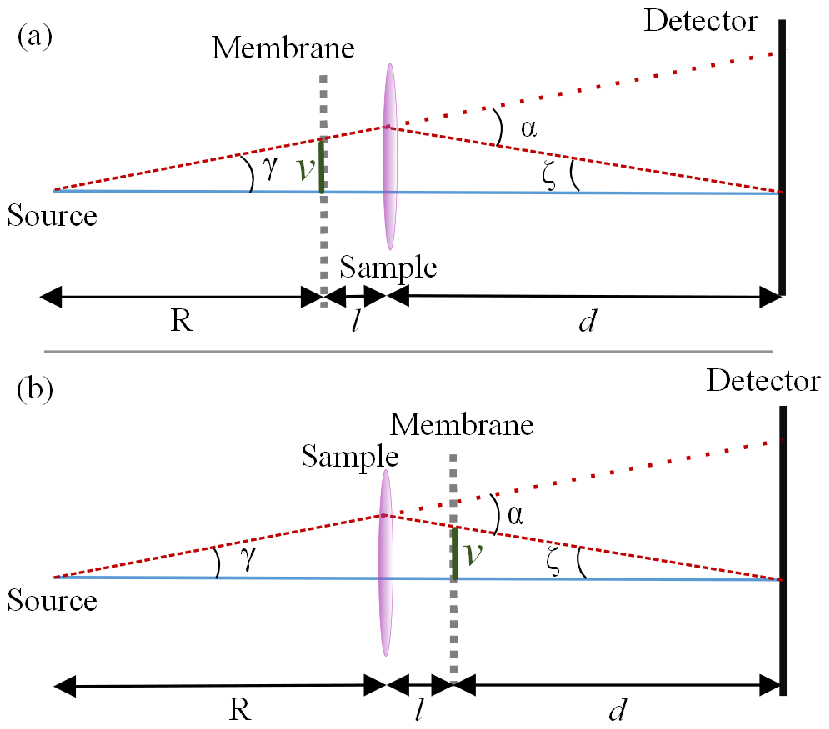}}
\caption{(Color online) The two possible geometries for the technique with the membrane placed either (a) upstream or (b) downstream the investigated object. The refraction angle of the (dashed) red ray passing through the sample is noted $\alpha$ and the (plain) blue ray shows the ray falling in the same detector pixel when no sample is in the beam.\label{fig:MembranePos}}
\end{figure}

Well known recipes based on correlation coefficient curve-fitting permit to calculate $\mathbf{v}$ with a fraction of a step accuracy \cite{pan2009,pan2006}. Then, from the angles $\alpha$, $\gamma$ and $\zeta$ defined in Fig.~\ref{fig:MembranePos} and for the two geometries (a) or (b), we have $\alpha=\gamma+\zeta$ and so by calculations:
\begin{equation}
\alpha\approx \left(\frac{R+d+l}{R}\right)\frac{ s_{s}}{d}\mathbf{v}
\end{equation}
which is valid under the small angle approximation encountered with X-rays. For a collimated beam, $\Gamma = (R+d+l)/R$ reduces to 1 and $\alpha=\zeta$.

In the following we use the del operator noted $\nabla$. As the angle $\alpha$ relates to the differential wavefront gradient $\alpha = \nabla W$ itself linked to the phase gradient $\nabla \phi$ \cite{born2008}, we can write:
\begin{equation}
k\nabla W=\nabla \phi(\mathbf{r}) \approx k\frac{s_{s}}{d}\Gamma\mathbf{v}(\mathbf{r})\label{eq:nablaphi}
\end{equation}
where $k$ is the wavenumber equal to $2\pi/\lambda$, with $\lambda$ the photon wavelength.
Most often, data recorded through 2D scans is followed by projection of $\mathbf{v}$ in order to extract the two orthogonal transverse phase gradient components. One-dimensional scans with much less points have also proved to be efficient for recovering the 1D phase gradient \cite{berujonPRA}.
The absorption, or here the transmission image $T$, can be correctly calculated free from the flat field correction errors that occur in presence of a strongly dephasing object and/or of a high magnification projection geometry \cite{stockmar2013,hagemann2014}. Its formulae can be obtained by consideration of Fig.~\ref{fig:MembranePos}:
\begin{equation}
T(\mathbf{r}) = \frac{\mu \left( f(\mathbf{r}) \right) }{\mu \left( g(\mathbf{r}+\Gamma s_s\mathbf{v}) \right)}
\end{equation}
where $\mu$ denotes the mean operator over the variable $\mathbf{p}$.
Finally, the darkfield image $D_f$ can be reconstructed by considering the decrease of speckle contrast that occurs when introducing the sample in the beam. More precisely one has to calculate the ratio of the coefficient of variations of $f$ and $g$:
\begin{equation}
D_f(\mathbf{r}) = \frac{1}{T(\mathbf{r})}\frac{\sigma\left( f(\mathbf{r})\right)}{\sigma\left( g(\mathbf{r}+\Gamma s_s \mathbf{v})\right)}
\label{eq:df}
\end{equation}
with $\sigma$ the standard deviation operator over $\mathbf{p}$.

Using a blueberry as a sample, the Figure~\ref{fig:mytille}.(a-d) illustrates the various modes the technique offers. The set of two scans was recorded with $d = 850$ mm, $s_{s} = 3~\mu$m, each scan being composed of an array of $25\times25$ points. The collection of such large data scans was performed in an attempt to optimize the calculation of the higher moments of the scattering distribution that require more statistics, as we shall see further. Otherwise many less points would be needed. Figure~\ref{fig:mytille} presents the calculated transmission image in (a), in (b) and (c) the respectively horizontal and vertical wavefront gradients $\nabla W.\mathbf{e_{x/y}}$, in (d) the integrated phase image, and in (e) the darkfield image.

\section{Ultra small angle X-ray scattering distributions\label{sec:usmangles}}
\begin{figure}
\centerline{\includegraphics[width=8cm]{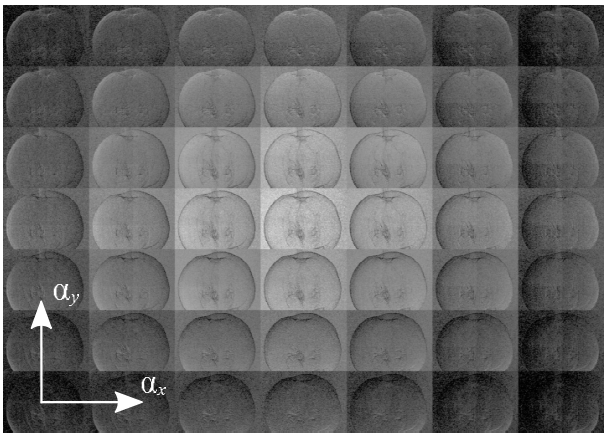}}
\caption{Array of the $5\times 5$ central scatter images organized by their angular vector $\alpha = s_{s}\mathbf{p}/d$ .\label{fig:decov}}
\end{figure}

Recent works demonstrated a way of recovering the scattering distribution of a sample in each detector pixel from raster scans. The concept was first introduced by Yashiro \emph{et al.} in \cite{yashiro2010} and \cite{yashiro2011} using a grating interferometer. Wherein, the scattering effect of the sample is treated as a local point spread function whose estimated width mirrors the local scattering strength. Later on, Modregger \emph{et al.} \cite{modregger2012,modregger2014} also used a grating interferometer but, this time, combined with numerical deconvolution to be able to extract the higher moments of the scattering distribution. In this section, we introduce by equivalence the recovery of the ultra-small angle X-ray scattering distribution using near field speckle scans similar to the ones recorded and used in the previous section.

Let us consider the light distribution $f$ in the case of a sample present in the X-ray beam path. Upon propagation, and for a given pixel position $\mathbf{r}$, we can consider $f$ as the effect of the sample optical transfer function $o(\mathbf{r},\mathbf{p})$ convoluted with the input reference signal $g$ \cite{modregger2012,berujonPRA}:
\begin{equation}
f(\mathbf{r},\mathbf{p}) = \int o(\mathbf{r},\mathbf{p}-\mathbf{u})g(\mathbf{r},\mathbf{p})d\mathbf{u}
\end{equation}

Henceforth, for each pair of patterns (with and without sample)	 $(f,g)(\mathbf{r})$, a deconvolution procedure can be applied to retrieve the distribution $o$. For this purpose the iterative Richardson-Lucy deconvolution \cite{richardson1972,lucy1974,biggs1997} algorithm is convenient because it is robust and its input does not require any assumption on the system noise \cite{scattarella2013}. Such processing was applied to the blueberry sample data of Sec.~\ref{sec:backgr} with 100 iterations used in the Richardson-Lucy deconvolution for each pixel. An illustration of the scattering distribution $o$ resulting from the deconvolution of the data of Fig.~\ref{fig:NormSketch}.(b) and (c) is displayed in (d).

Geometrically, the position $\mathbf{p}$ corresponds to the angular coordinate $\alpha = \Gamma\frac{s_{s}}{d}\mathbf{p}$ (see Fig.~\ref{fig:MembranePos}). By projection in this angular variable coordinate system, a scattering image $o(\mathbf{\alpha})$ can be constructed for each angular vector. Figure~\ref{fig:decov} shows an image array of the blueberry sample organized as a function of the scattering angular coordinates. Given a scan step of $s_{s} = 3~\mu$m, the angular step size from image to image was of 3.5 $\mu$rad.

Moments of the scattering distribution contain physical information about the sample features and correspond, as we shall see for the first of them, to the traditional contrast modalities, i.e. the absorption, phase and scattering signals. The central moments $M_{mn}$ of order $(m,n), m+n >1$ of $o(\mathbf{r})$ are calculated within the meaning of angular probability density functions:
\begin{equation}
M_{mn}(\mathbf{r}) = \frac{1}{M_{00}}\left(\frac{s_{s}}{d}\right)^{(m+n)}\int (X-\overline{X_{\mathbf{r}}})^m (Y-\overline{Y_{\mathbf{r}}})^n o(\mathbf{r},\mathbf{p})d\mathbf{p}
\end{equation}
where:
\begin{equation}
M_{00} = \int o(\mathbf{r},\mathbf{p})d\mathbf{p}
\end{equation}
and $(\overline{X}_{\mathbf{r}},\overline{Y}_{\mathbf{r}})$ are the centroid positions of $o(\mathbf{r},\mathbf{p})$:
\begin{equation}
\overline{X}_{\mathbf{r}} = \frac{\int X o(\mathbf{r},\mathbf{p})d\mathbf{p}}{M_{00}}~~,~~\overline{Y}_{\mathbf{r}} = \frac{\int Y o(\mathbf{r},\mathbf{p})d\mathbf{p}}{M_{00}}
\end{equation}

We have then $\mathbf{v}(\mathbf{r}) = (\overline{X}_{\mathbf{r}},\overline{Y}_{\mathbf{r}})$ by definition of $\mathbf{v}$, from which we can calculate $\nabla \phi$ using Eq.~\ref{eq:nablaphi}. Furthermore, the two second normalized moments $\sqrt{M_{02}}$ and $\sqrt{M_{20}}$ mirror the angular characteristic scattering width in the two transverse directions. Higher moments provide complementary information on the scattering distribution, with, for instance, the third and fourth moments related to its skewness and kurtosis.
The wavefront differential gradients $\nabla W.\mathbf{e_{x/y}}$ for the blueberry sample obtained using the scattering distribution centroid are shown in Fig.~\ref{fig:mytille}.(f) and (g). Even visually they are in good agreement with the equivalent maps (b) and (c) calculated with the constant step method presented in the previous section. In this same figure, the second moments $\sqrt{M_{02}}$ and $\sqrt{M_{20}}$ are shown in (i) and (j) and the higher moments $\sqrt[3]{M_{30}}$, $\sqrt[3]{M_{03}}$ and $\sqrt[4]{M_{40}}$ drawn in (k,l) and (h).

\section{Scans with random steps\label{sec:randstep}}

In this section, we present a processing method to recover the sample-induced 2D phase gradients from images acquired using random-step scans, i.e. scans where the images are recorded with the scattering membrane moved at random transverse positions $(X,Y)$ in the plan $(\mathbf{e_x},\mathbf{e_y})$, with no particular translation distance from one image to the next. However, it is crucial that scans performed with and without the sample have their images taken with the scattering membrane located at the very same positions for each corresponding step.

\begin{figure}
\centerline{\includegraphics[width=8cm]{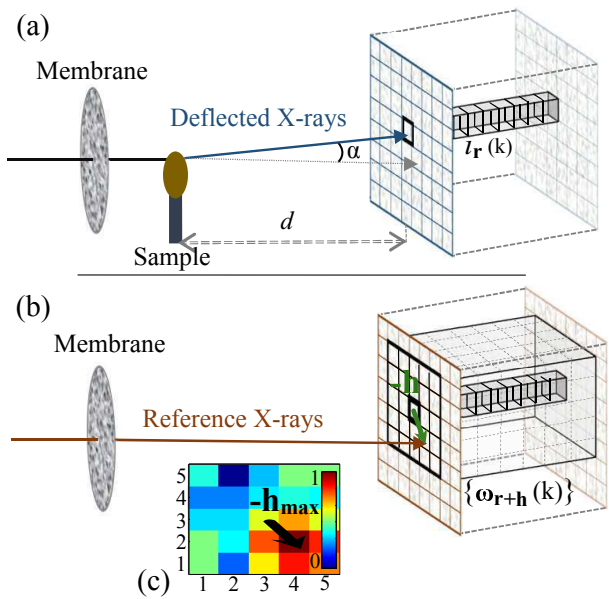}}
\caption{(Color online) Geometry of the concept. (a) Data vector $\mathbf{i}_\mathbf{r}$ built from $N$ images with the sample in the beam. (b) Set of reference vectors $\{\omega_{\mathbf{r+h}}\}$ in the reference data. (c) Correlation factor map of $\rho(\mathbf{i}_\mathbf{r},\mathbf{\omega}_\mathbf{r+h})$. \label{fig:randstep}}
\end{figure}

Here, we will change slightly the notation, in order not to confuse the reader with the previous processing method. We note $i(\mathbf{r},n)$ the intensity collected at the position $\mathbf{r}$ of the $n^{th}$ image of a scan containing $N$ images when the sample is inserted into the beam. Then, writing $i_{\mathbf{r},n} = i(\mathbf{r},n)$, we have the vectors $\mathbf{i}_\mathbf{r} = (i_{\mathbf{r},1},...,i_{\mathbf{r},n},...,i_{\mathbf{r},N)}$ built from $N$ realizations of the random variables $I_\mathbf{r}$, induced by the variable position of the membrane. With equivalent notations, we also build the reference vectors $\mathbf{\omega}_\mathbf{r}$, from $N$ realizations of $W_\mathbf{r}$ obtained in absence of sample.

The idea is here again to use correlation calculations between the speckle distribution in the various pixels to retrieve the beam wavefront derivative through the calculation of the local light deflection angles $\alpha(\mathbf{r})$ as sketched in Fig.~\ref{fig:randstep}. So, we use the Pearson correlation coefficient $\rho_{\mathbf{r}}$ generally defined by:
\begin{equation}
\rho(I_{\mathbf{r}},W_{\mathbf{r\prime}}) = \frac{E\left[(I_\mathbf{r} - \mu({I_r}) )(W_\mathbf{r\prime} - \mu (W_\mathbf{r\prime}) )\right]}{\sigma (I_{\mathbf{r}}) \sigma ({W_\mathbf{r\prime}}) }
\label{eq:correq}
\end{equation}
where $E$ denotes the expected value. The factor $\rho$ is the normalized covariance of the pixel signal distributions. It is used as figure of merit to evaluate the level of similarity between the data acquired in the different pixels of the two scans.
For the vectors $\mathbf{i}$ and $\mathbf{\omega}$, this operation can also be seen as a scalar product. Thus, we calculate an estimate of each projection vector of $\mathbf{i}_\mathbf{r}$ onto vecrors $\mathbf{\omega}_\mathbf{r+h}$ located at a small distance $\mathbf{h}$ from $\mathbf{r}$ in the reference scans:
\begin{equation}
\rho(\mathbf{i}_\mathbf{r},\mathbf{\omega}_\mathbf{r+h}) = \frac{\sum\limits_k^N{(i_{\mathbf{r},k} - \overline{i_\mathbf{r}})(\omega_{\mathbf{r+h},k} - \overline{\omega_\mathbf{r+h}})}}{ \sqrt {\sum\limits_k^N{(i_{\mathbf{r},k} - \overline{i_\mathbf{r}})^2}\sum\limits_{k}^{N} {(\omega_{\mathbf{r+h},k} - \overline{\omega_\mathbf{r+h}})^2 }}}\label{eq:correstimate}
\end{equation}
where $\overline{i_\mathbf{r}}$ and $\overline{\omega_\mathbf{r}}$ denote the mean values of the measurement vectors.

For all $\mathbf{i_r}$, and with the correlation map $\rho(\mathbf{i}_\mathbf{r},\mathbf{\omega}_\mathbf{r+h})$ (see Fig. \ref{fig:randstep}.(c)), we can locate the vector position of $\mathbf{\omega_{\mathbf{r+h_{max}}}}$ rendering maximum correlation:
\begin{equation}
\mathbf{h}_{max} = \argmax_{\mathbf{h}}\rho(\mathbf{i}_\mathbf{r},\mathbf{\omega}_\mathbf{r+h})
\end{equation}

Here also the vector $\mathbf{h_{max}}$ is determined with a subpixel accuracy by numerical interpolation over the neighboring pixels \cite{tian1986,pan2006}.
Following the law of light propagation in homogeneous media, the phase gradient $\nabla\phi$ is recovered by:
\begin{equation}
\nabla\phi (\mathbf{r}) = k\frac{s_{p} }{d}\mathbf{h}_{max}(\mathbf{r})
\end{equation}
where $s_p$ is the detector pixel size.
The corrected transmission signal is equal to:
\begin{equation}
T_{\mathbf{r}} = \frac{\mu(I_{\mathbf{r}})} {\mu(W_{\mathbf{r+h_{max}(\mathbf{r})}})}
\end{equation}
and is valid for any geometry.
With this processing method the darkfield signal writes:
\begin{equation}
Df = \frac{1}{T_{\mathbf{r}}} \frac{\sigma (I_{\mathbf{r}})}{\sigma (W_{\mathbf{r+h_{max}(\mathbf{r})}})}
\end{equation}
which is fully equivalent to Eq.~\ref{eq:df}.
This method can be seen as an over-sampling version of the X-ray speckle tracking technique. Indeed, the cross-correlation operation is operated between data collected in different pixels as performed with speckle tracking \cite{berujonPRL}. Moreover, the refraction angle $\alpha$ is here calculated at the sample position (see Fig.~\ref{fig:randstep}) and not at the detector plane level as in the case of the method of Sec.~\ref{sec:backgr} with the angle $\zeta$.
\begin{figure}
\centerline{\includegraphics[width=8cm]{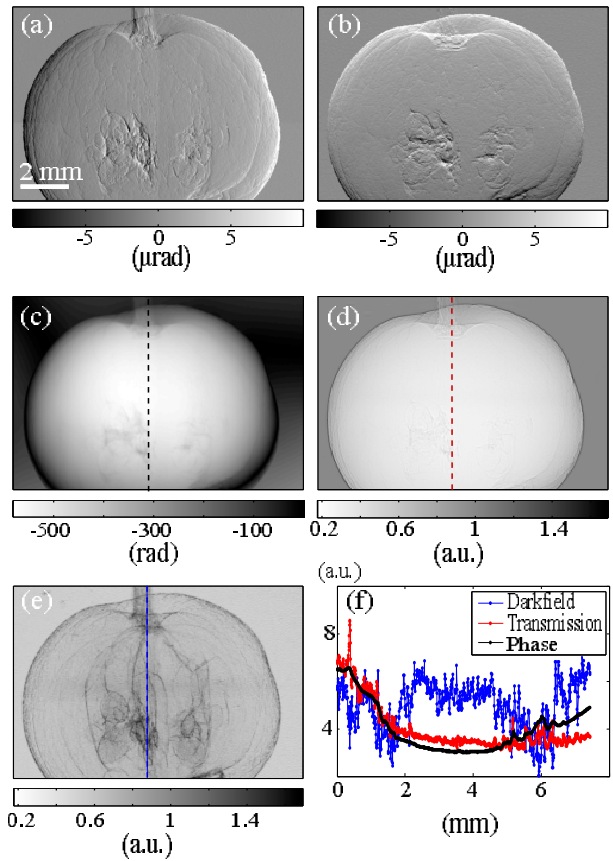}}
\caption{(Color online) Random step processing scheme. (a) Horizontal and (b) vertical differential wavefront gradients. (c) Phase (d) transmission and (e) scattering reconstruction images. (f) Intensity plot along the lines marked in (c-e).\label{fig:randexp}}
\end{figure}

An experimental demonstration of the processing scheme is provided in Fig.~\ref{fig:randexp}, using the same blueberry sample as before, for comparison purpose. A pair of 25-image scans was acquired by applying displacements to the membrane with steps larger than 20 $\mu$m using a micrometer precision linear positioning motor. Fig.~\ref{fig:randexp}.(a) and (b) are respectively showing the horizontal and vertical transverse differential wavefront gradients. The results obtained are in nearly perfect agreement with the one calculated with the constant step size method presented in Sec. \ref{sec:backgr}. Figure~\ref{fig:randexp} shows the reconstructed phase shift of the sample (c), the transmission image (d) and the darkfield image (e). Figure~\ref{fig:randexp} (f) displays a plot of cuts along the marked lines of the three previous images normalized to an arbitrary scale. One can observe the way the different contrasts vary depending on the image modality.

This processing scheme is particularly interesting as it allows to recover the 2D phase and scattering information for every pixel, as does the scheme with constant step, although requiring a lower number of images. With the constant-step method, $\mathcal{O}(N^2)$ images are necessary to build the 4D matrix $f(\mathbf{r},\mathbf{p})$ whilst only $\mathcal{O}(N)$ are required in this method for 2D phase sensing.

\section{Discussion}
\subsection{Sensitivity}

The sensitivity of the speckle scanning technique differs depending on the processing method. When applying the algorithm presented in Sec.~\ref{sec:backgr} on data acquired with a scan of constant step, the accuracy $\sigma(\alpha)$ on the wavefront gradient is:
\begin{equation}
\sigma_1(\alpha) = \frac{s_{s}}{d}\Gamma\sigma(\mathbf{v_r})
\end{equation}

This can be compared to the processing method introduced in Sec.~\ref{sec:randstep}, for instance when using the same data:
\begin{equation}
\sigma_2(\alpha) = \frac{s_{p}}{d}\sigma(\mathbf{h_r})
\end{equation}

From these equations, with a similar experimental setup, it is the constant-step scanning method that delivers the highest sensitivity. As a matter of fact, considering an equivalent sub-pixel/step accuracy for the peak finder algorithm \cite{pan2009} in both processing methods, we have $\sigma(\mathbf{v_r}) \approx \sigma(\mathbf{h_r})$ and we can set $s_{s}<s_{p} $ to diminish $\sigma_1$. When working with a collimated X-ray beam, the gain in sensitivity gets rapidly limited by the minimum step length necessary to get a sufficient intensity variation within a pixel from one step to the next \cite{pan2006,pan2008}. On the contrary, with a magnification projection geometry setup, the gain can be tremendous since a tiny displacement of the membrane located near the X-ray beam focal point can generate a much larger displacement of the speckle pattern at the detector level. In this case, when generating a sufficient speckle statistic variation for the correlation algorithm to be efficient and accurate, the angular resolution may reach the nanoradian scale \cite{berujon2014}.

The convergence of the distribution estimators in Eq.~\ref{eq:correstimate} scales with $1/\sqrt{N}$ \cite{riley2006}. This implies that only the few first images are necessary to reach a suitable accuracy since $\sqrt{N}\sigma(\mathbf{h})\rightarrow 0$ and $\sqrt{N}\sigma(\mathbf{v})\rightarrow 0$ as $N \rightarrow \infty$.

From Eq.~\ref{eq:correq}, considering two identical scans or areas of $\mathbf{r}$ where the X-ray light is neither refracted nor scattered by any sample, the distributions $(I_{\mathbf{r}},W_{\mathbf{r}})$ should, by definition, render a theoretical value $\rho(I_{\mathbf{r}},W_{\mathbf{r}})= 1$. For our experimental data, when using the method of Sec.~\ref{sec:randstep}, the deviation from unity of $\rho$ over an area of $150\times 150$ pixels amounts to $\sigma(\alpha) = 0.8~\mu$rad for a scan of 25 images. When more than 150 images are used it tends to a limit of $\sigma(\alpha) = 0.35~\mu$rad. This last figure is comparable to the one calculated with the two constant-step methods presented in Sec.~\ref{sec:backgr} and ~\ref{sec:usmangles}, suggesting that the discrepancy may be due to monochromattor instabilities, more felt in the vertical direction.

\subsection{Tracking orientation}

Section \ref{sec:randstep} explains and illustrates the way the refraction angle is recovered by examining the vector positions in the original reference scan data. This subsection aims at illustrating the importance of operating along this line rather than tracking the reference scan vectors in the data array collected when the sample is present in the beam. The problem raised here is due to the creation of optical vortices in the propagated beam when going through the sample discontinuities \cite{paganin2006}.
\begin{figure}
\centerline{\includegraphics[width=8cm]{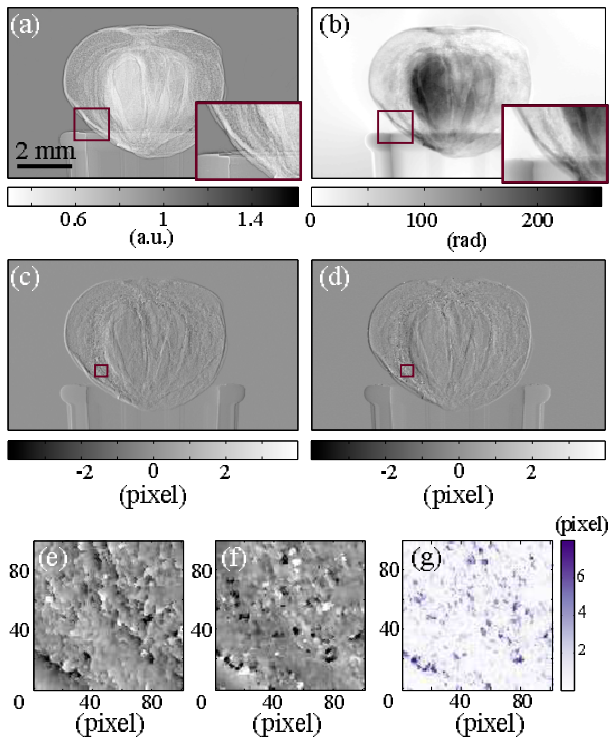}}
\caption{(Color online) (a-b) Correct transmission and phase images of a juniper cone. (c) Horizontal phase gradient calculated searching the location of vectors from the sample scan across the reference scan. (d) Output of the reciprocal calculation, searching reference vector locations across the sample scan. (e-f) Zoom in the small subset drawn on the above images. (f) Absolute difference between the values of two magnified image subsets: $\Delta = (e)-(f)$.\label{fig:optvort}}
\end{figure}

Figure~\ref{fig:optvort} presents an experimental data set for a juniper berry sample imaged with X-rays. That sample exhibits strong scattering features and turbid phase shift. Figure~\ref{fig:optvort}.(a) and (b) are successfully displaying the transmission and phase contrast images, respectively. Figure~\ref{fig:optvort}.(c) presents the horizontal speckle displacement vectors calculated picking vectors within the sample scan and searching for their counterpart vectors in the reference scan. Conversely, (d) corresponds to the tracking of vectors from the reference scan across the sample scan data. Figure \ref{fig:optvort}.(e-f) are zooming in the square subsets of (c) and (d), while (g) shows the absolute intensity difference between these two subsets. The significant difference observed between the two calculated images results from the incapacity of the algorithm to seize reference vectors folded up in optical vortices upon propagation. As these vectors add up at singularity points they generate new distributions $I_{\mathbf{r}}$ \cite{paganin2006} that are no longer correlated to any reference vector. Obviously, as the detector pixel size and the speckle grains themselves are getting smaller, or as the propagation distance increases, the effect becomes more pronounced.

\section{Conclusion}
Several advanced processing schemes were demonstrated within the context of a new X-ray speckle scanning imaging technique. Starting from the presently available scanning method with a constant step, we proposed ways of building error free transmission images and of accessing the sample scattering distribution, pixel by pixel, through a numerical deconvolution of the scans data. The various processing schemes were demonstrated with experimental data, assessing thereby the excellence of the method.
The recovery of the sample scattering distribution is of interest in various fields of material science as well as in biological imaging where the orientations of sub-pixel structures are responsible for the functional behavior of the systems.
In addition, a new processing scheme with relaxed constraint on the step size was proposed. It can be seen as an over sampled version of the speckle tracking technique with a resolution enhanced up to the one defined by the detector. The speckle-scanning imaging technique based on a random step scheme reduces dramatically the number of acquisitions as compared to the previously available schemes. This point is often essential in the case of biological imaging, where the photon efficiency permits to reduce the dose delivered to the sample. Future work will permit to push and test the limit of the method in this regard. The proposed approach will also promote the development of the speckle imaging technique with laboratory low brilliance sources.

\begin{acknowledgments}
The authors wish to thank the ESRF for the financial support, S. Da Cunha for assistance when setting up experiments and J. Susini for the encouragements and support.
\end{acknowledgments}


%

\end{document}